\documentstyle[12pt]{article}
\newlength{\extraspace}
\newlength{\extraspaces}
\newcommand{\be}{\begin{equation}
\addtolength{\abovedisplayskip}{\extraspaces}
\addtolength{\belowdisplayskip}{\extraspaces}
\addtolength{\abovedisplayshortskip}{\extraspace}
\addtolength{\belowdisplayshortskip}{\extraspace}}
\newcommand{\ee}{\end{equation}}
\newcommand{\ba}{\begin{eqnarray}
\addtolength{\abovedisplayskip}{\extraspaces}
\addtolength{\belowdisplayskip}{\extraspaces}
\addtolength{\abovedisplayshortskip}{\extraspace}
\addtolength{\belowdisplayshortskip}{\extraspace}}
\newcommand{\ea}{\end{eqnarray}}
\newcommand{\nonu}{\nonumber \\[.5mm]}
\newcommand{\A}{&\!\!\!}

\begin{document}
\addtolength{\baselineskip}{.7mm}
\begin{flushright}
STUPP-96-143 \\ February, 1996
\end{flushright}
\vspace{.6cm}
\begin{center}
{\large{\bf{Consistency of matter field equations \\[2mm]
              in Ashtekar formulation}}} \\[20mm]
{\sc Motomu Tsuda and Takeshi Shirafuji} \\[12mm]
{\it Physics Department, Saitama University \\[2mm]
Urawa, Saitama 338, Japan} \\[20mm]
{\bf Abstract}\\[10mm]
{\parbox{13cm}{\hspace{5mm}
The chiral Lagrangian including more than two spin-3/2 
fields becomes complex in general 
when the self-dual connection satisfies 
its equation of motion, 
and the imaginary part of the chiral Lagrangian 
gives the additional equation for spin-3/2 fields 
which gives rise to the inconsistency. 
This inconsistency can be removed 
by taking the tetrad to be complex 
and defining the total Lagrangian 
which is the sum of the chiral Lagrangian 
and its complex conjugate. 
It is possible to establish the right (left)-handed 
supersymmetry in its total Lagrangian 
as in the case of the chiral Lagrangian. 
We also comment on the canonical formulation 
of the total Lagrangian.}} 
\end{center}
\vfill

\newpage
%

The Ashtekar refomulation of canonical gravity \cite{AA}-\cite{AAL} 
can be derived from the chiral Lagrangian in which a tetrad 
and a complex self-dual connection are regarded 
as independent variables. 
Although the chiral Lagrangian is complex, 
its imaginary part vanishes in the source-free case 
when the tetrad is assumed to be real 
and the equation of self-dual connection is satisfied: 
The chiral Lagrangian then reduces 
to the Einstein-Hilbert Lagrangian. 
If the matter terms of integer spin fields exist, 
the chiral Lagrangian also becomes real 
as in the source-free case, 
because the Lagrangian of integer spin fields does not 
contain the self-dual connection. 
Moreover, it has been shown for spin-1/2 fields \cite{AAL,ART} 
and simple supergravity \cite{AAN,JJ,CDJ} that the imaginary 
part of the chiral Lagrangian vanishes 
(up to a total divergence term). 
But in general, the chiral Lagrangian becomes complex 
for more than two spin-3/2 fields, 
and its imaginary part gives the additional equation 
for spin-3/2 fields which gives rise to the inconsistency \cite{TSX}. 
We suggest in this letter one of the possible ways to evade 
the inconsistency of matter field equations. 

We consider a spacetime manifold with a metric $g_{\mu \nu}$ 
constructed from a tetrad $e_{\mu}^i$ 
via $g_{\mu \nu} = e^i_{\mu} e^j_{\nu} \eta_{ij}$, 
and denote a self-dual connection 
by $A^{(+)}_{ij \mu} = A^{(+)}_{[ij] \mu}$
which satisfy 
\be
A^{(+) \ast}_{ij \mu} := {1 \over 2} {\epsilon_{ij}} \! ^{kl} A^{(+)}_{kl \mu} 
                   = i A^{(+)}_{ij \mu}, 
\ee
where the $\ast$ operation means the duality operation 
in Lorentz indices.
\footnote{\ Greek letters $\mu, \nu, \cdots$ are 
space-time indices, and Latin letters {\it i, j,} $\cdots$ 
are local Lorentz indices. 
We denote the Minkowski metric 
by $\eta_{ij} =$ diag$(-1, +1, +1, +1)$. 
The totally antisymmetric tensor $\epsilon_{ijkl}$ 
is normalized as $\epsilon_{0123} = +1$. 
The antisymmetrization of a tensor is denoted 
by $A_{[ij]} := (1/2)(A_{ij} - A_{ji})$.} 
In order to see the inconsistency of matter field equations, 
let us take $N$-(Majorana) Rarita-Schwinger fields 
$\psi_{\mu}$ \ ($:= \psi_{\mu}^I$ with $I$ running from $1$ to $N$) 
as example. 
The chiral Lagrangian density ${\cal L}^{(+)}$ 
including $N$-(Majorana) Rarita-Schwinger fields is 
\be
{\cal L}^{(+)} = {\cal L}^{(+)}_G + {\cal L}^{(+)}_{RS}. 
\label{L+}
\ee
Here ${\cal L}^{(+)}_G$ is the chiral gravitational 
Lagrangian density constructed from the tetrad 
and the self-dual connection: 
\be
{\cal L}^{(+)}_G = -{i \over 2} e \ \epsilon^{\mu \nu \rho \sigma} 
   e_{\mu}^i e_{\nu}^j R^{(+)}_{ij \rho \sigma}, 
\label{LG+}
\ee
where the unit with $8 \pi G = c = 1$ is used, 
$e$ denotes ${\rm det}(e^i_{\mu})$ 
and the curvature of self-dual connection 
${R^{(+)ij}}_{\mu \nu}$ is 
\be
{R^{(+)ij}}_{\mu \nu} := 2(\partial_{[\mu} {A^{(+)ij}}_{\nu]} 
             + {A^{(+)i}}_{k [\mu} {A^{(+)kj}}_{\nu]}). 
\label{curv+}
\ee
On the other hand, ${\cal L}^{(+)}_{RS}$ is the chiral 
Lagrangian density of $N$-(Majorana) Rarita-Schwinger fields, 
which is obtained by assuming that the fields $\psi_{\mu}$ 
are minimally coupled to gravity and that the Lagrangian 
is described by using only the self-dual connection: 
\be
{\cal L}^{(+)}_{RS} = - e \ \epsilon^{\mu \nu \rho \sigma} 
                     \overline \psi_{R \mu} \gamma_\rho 
                     D^{(+)}_\sigma \psi_{R \nu}, \label{LRSE+}
\ee
where $\psi_R = (1/2)(1 + \gamma_5) \psi$ 
and $D^{(+)}_\mu$ denotes the covariant derivative 
with respect to $A^{(+)}_{ij \mu}$: 
\be
D^{(+)}_\mu := \partial_\mu + {i \over 2} A^{(+)}_{ij \mu} S^{ij} 
\ee
with $S_{ij}$ standing for the Lorentz generator. 

When the self-dual connection satisfies its equation of motion, 
the imaginary part of ${\cal L}^{(+)}$ 
vanishes for $N=1$, while it does not vanish for $N \ge 2$. 
To see this, vary the ${\cal L}^{(+)}$ of (\ref{L+}) 
with respect to $A^{(+)}_{ij \mu}$ to obtain 
\be
\epsilon^{\mu \nu \rho \sigma} D^{(+)}_{\nu} {H^{(+)ij}} \! _{\rho \sigma} 
    = {i \over 2} \epsilon^{\mu \nu \rho \sigma} 
      \left( e_{\nu}^{[i} \ \overline \psi_{\rho} \gamma^{j]} \psi_{\sigma} 
            - {i \over 2} {\epsilon^{ij}} \! _{kl} 
           e_{\nu}^k \ \overline \psi_{\rho} \gamma^l \psi_{\sigma} \right) 
\label{eqA+}
\ee
with ${H^{(+)ij}} \! _{\mu \nu}$ being defined 
by ${H^{(+)ij}} \! _{\mu \nu} := e_{[\mu}^i e_{\nu]}^j 
- (i/2) {\epsilon^{ij}} \! _{kl} e_{\mu}^k e_{\nu}^l$, 
and $D^{(+)}_{\mu} e_{\nu}^i = \partial_{\mu} e_{\nu}^i 
+ {A^{(+)i}}_{j \mu} e_{\nu}^j$. 
Equation (\ref{eqA+}) can be solved 
with respect to $A^{(+)}_{ij \mu}$ as 
\be
A^{(+)}_{ij \mu} = A^{(+)}_{ij \mu}(e) + K^{(+)}_{ij \mu}, 
\label{solA+}
\ee
where $A^{(+)}_{ij \mu}(e)$ is the self-dual part 
of the Ricci rotation coefficients $A_{ij \mu}(e)$, 
while $K^{(+)}_{ij \mu}$ is that of $K_{ij \mu}$ given by 
\be
K_{ij \mu} := {i \over 4} (e^{\rho}_i e^{\sigma}_j e_{\mu}^k 
             \ \overline \psi_{\rho} \gamma_k \psi_{\sigma} 
    + e^{\rho}_i 
             \ \overline \psi_{\rho} \gamma_j \psi_{\mu} 
    - e^{\rho}_j 
             \ \overline \psi_{\rho} \gamma_i \psi_{\mu}). 
\ee
If the tetrad is real, substituing the solution of (\ref{solA+}) 
back into the ${\cal L}^{(+)}$ of (\ref{L+}) gives 
the imaginary part of ${\cal L}^{(+)}$ as follows: 
\be
{\rm Im}{\cal L}^{(+)} = - {1 \over 32} 
              \sum_{I,J} \epsilon^{\mu \nu \rho \sigma} 
             (\overline \psi_{\mu}^I \gamma_i \psi_{\nu}^I) 
             (\overline \psi_{\rho}^J \gamma^i \psi_{\sigma}^J), 
\label{ImL+}
\ee
where we denote the indices $I,J$ explicitly. 
Now the Im${\cal L}^{(+)}$ of (\ref{ImL+}) does not vanish 
except for $N = 1$. 
This non-vanishing Im${\cal L}^{(+)}$ yields the additional equation 
$\partial {\rm Im}{\cal L}^{(+)}/\partial \psi_{\mu}^I = 0$ 
in general, 
besides the Einstein equation and the Rarita-Schwinger equation 
obtained by taking variation of Re${\cal L}^{(+)}$ 
with respect to the tetrad 
and the (Majorana) Rarita-Schwinger fields. 
Therefore the difficulty of overdetermination 
for $\psi_{\mu}^I$ arises 
and the matter field equations generally become inconsistent.
\footnote{\ This inconsistency cannot be removed only by taking the tetrad 
to be complex in ${\cal L}^{(+)}$, 
because all terms in ${\cal L}^{(+)}$ 
(up to the term of (\ref{ImL+})) merely becomes to be complex, 
and so the additional equations of $\psi_{\mu}^I$ also appear 
as in the case of the real tetrad.} 

In order to remove this inconsistency of matter field equations, 
one could take ${\cal L}^{(+)}$ to be analytic for all fields; 
namely, the complex tetrad, the self-dual connection, 
the independent complex spinor fields 
$\psi_R$ and $\overline \psi_R$. 
If we follow this procedure and take the tetrad 
to be complex in the Lagrangian of spin-1/2 fields, 
we have to take $\psi$ and $\overline \psi$ to be independent 
in order to keep the matter field equations consistent. 
But $\psi$ and $\overline \psi$ cannot be independent 
in QED or QCD, because if they are independent, 
the electric current, for example, will become complex. 

So we adopt the following procedure as a simplest way 
to eliminate the term of (\ref{ImL+}): 
Namely, we add the complex conjugate of the chiral Lagrangian 
density, $\overline{{\cal L}^{(+)}}$, 
to the Lagrangian density ${\cal L}^{(+)}$, 
and define a total Lagrangian density ${\cal L}^{tot}$ by 
\be
{\cal L}^{tot} := {\cal L}^{(+)} + \overline{{\cal L}^{(+)}}. 
\label{Ltot}
\ee
Here we take the tetrad to be complex, 
because if the tetrad is real, ${\cal L}^{tot}$ reduces 
to the Lagrangian for ordinary Einstein gravity. 
The basic variables in ${\cal L}^{(+)}$ 
are the complex tetrad, the self-dual connection, 
the spinor fields $\psi$ and its Dirac conjugate $\overline \psi$, 
while the basic variables in the ${\cal L}^{tot}$ 
of (\ref{Ltot}) are the set of $Q = (e, A^{(+)}, \psi)$ 
and their complex conjugate $\overline Q$. 

>From the definition of (\ref{Ltot}), 
the total gravitational Lagrangian density, 
${\cal L}^{tot}_G$, is written by 
\be
{\cal L}^{tot}_G := -{i \over 2} e \ \epsilon^{\mu \nu \rho \sigma} 
   e_{\mu}^i e_{\nu}^j R^{(+)}_{ij \rho \sigma} + {\rm c.c.}, 
\label{LGtot}
\ee
where ``c.c.'' means ``the complex conjugate of the preceding term''. 
It can be shown that ${\cal L}^{tot}_G$ describes 
the sum of complex general relativity and its complex conjugate 
when the self-dual connection (and its complex conjugate) 
satisfies its equations of motion. 
In the case of $N$-(Majorana) Rarita-Schwinger fields coupling, 
since the term of (\ref{ImL+}) does not contain the tetrad, 
it cancels with its complex conjugate in ${\cal L}^{tot}$. 
Accordingly, the inconsistency of matter field equations disappears. 
Furthermore we can show that ${\cal L}^{tot}$ describes 
the sum of complex Einstein gravity with four-fermion contact terms 
and its complex conjugate, 
after the equation of self-dual connection 
(and its complex conjugate) is solved. 

The right (left)-handed supersymmetry 
in the chiral Lagrangian for $N = 1$ supergravity 
has been established \cite{AAN,JJ,CDJ}. 
In the same way, it is possible to establish 
right (left)-handed supersymmetry in ${\cal L}^{tot}$ for $N = 1$, 
if we now take $\psi_{R \mu}$ and $\overline \psi_{R \mu}$ 
in (\ref{LRSE+}) to be independent 
and define the total Lagrangian density 
of (Majorana) Rarita-Schwinger fields 
by ${\cal L}^{tot}_{RS} := {\cal L}^{(+)}_{RS} 
+ \overline{{\cal L}^{(+)}_{RS}}$; 
namely 
\be
{\cal L}^{tot}_{RS} = - e \ \epsilon^{\mu \nu \rho \sigma} 
                     \overline {\tilde \psi}_{R \mu} \gamma_\rho 
                     D^{(+)}_\sigma \psi_{R \nu} + {\rm c.c.} 
\label{LRSEtot}
\ee
with $\overline {\tilde \psi}_{R \mu}$ and $\psi_{R \mu}$ 
being independent. 
Note that in special relativistic limit, the ${\cal L}^{tot}_{RS}$ 
of (\ref{LRSEtot}) becomes 
\be
L^{tot}_{RS} = \epsilon^{\mu \nu \rho \sigma} 
               \overline {\tilde \psi}_{\mu} \gamma_5 \gamma_\rho 
               \partial_\sigma \psi_{\nu}, \label{LRStots}
\ee
which can be diagonalized as 
\be
L^{tot}_{RS} = \epsilon^{\mu \nu \rho \sigma} 
              (\overline \psi_{\mu}^1 \gamma_5 \gamma_\rho 
               \partial_\sigma \psi_{\nu}^1 
             - \overline \psi_{\mu}^2 \gamma_5 \gamma_\rho 
               \partial_\sigma \psi_{\nu}^2), 
\label{LRStots2}
\ee
with $\psi_{\mu}^1 := (1/2)(\psi_{\mu} + {\tilde \psi}_{\mu})$ 
and $\psi_{\mu}^2 := (1/2)(\psi_{\mu} - {\tilde \psi}_{\mu})$. 
The minus sign in (\ref{LRStots2}) means 
the appearance of negative energy states. 

The total Lagrangian density ${\cal L}^{tot}$, 
which is the sum of (\ref{LGtot}) and (\ref{LRSEtot}), 
is invariant 
under the right-handed supersymmetry transformation 
generated by anticommuting spinor parameter $\alpha$, 
\ba
\begin{array}{ll}
\delta_R \psi_{R \mu} = 2 D^{(+)}_{\mu} \alpha_R 
& \ \ \delta_R \overline{\tilde \psi}_{R \mu} = 0, \vspace{1mm} \\
\delta_R \psi_{L \mu} = 0 
& \ \ \delta_R \overline{\tilde \psi}_{L \mu} 
  = 2 \overline{\tilde D^{(-)}_{\mu} \tilde \alpha}_L, \vspace{1mm} \\
\delta_R e_{\mu}^i 
= -i \overline{\tilde \psi}_{R \mu} \gamma^i \alpha_R 
& \ \ \delta_R \overline{e_{\mu}^i} 
  = i \overline{\tilde \alpha}_L \gamma^i \psi_{L \mu}, 
\end{array}
\ea
and the left-handed supersymmetry transformation, 
\ba
\begin{array}{ll}
\delta_L \psi_{R \mu} = 0 
& \ \ \ \delta_L \overline{\tilde \psi}_{R \mu} 
  = 2 \tilde D^{(-)}_{\mu} \overline{\tilde \alpha}_R, \vspace{1mm} \\
\delta_L \psi_{L \mu} = 2 \overline{D^{(+)}_{\mu}} \alpha_L 
& \ \ \ \delta_L \overline {\tilde \psi}_{L \mu} = 0, \vspace{1mm} \\
\delta_L e_{\mu}^i 
= i \overline {\tilde \alpha}_R \gamma^i \psi_{R \mu} 
& \ \ \ \delta_L \overline {e_{\mu}^i} 
  = -i \overline {\tilde \psi}_{L \mu} \gamma^i \alpha_L, 
\end{array}
\ea
when we use the equation of self-dual connection 
(and its complex conjugate). 
Here $\tilde D^{(-)}_{\mu}$ denotes the covariant derivative 
with respect to antiself-dual connection, 
$\tilde A^{(-)}_{ij \mu}$, 
and we assume that $\tilde A^{(-)}_{ij \mu}$ is the solution 
derived from the Lagrangian, 
$\tilde {\cal L}^{(-)} + \overline{\tilde {\cal L}^{(-)}}$ 
with $\tilde {\cal L}^{(-)}$ being 
\be
\tilde {\cal L}^{(-)} = {i \over 2} e \ \epsilon^{\mu \nu \rho \sigma} 
   e_{\mu}^i e_{\nu}^j \tilde R^{(-)}_{ij \rho \sigma} 
                   + e \ \epsilon^{\mu \nu \rho \sigma} 
                     \overline{\tilde \psi}_{L \mu} \gamma_\rho 
                     \tilde D^{(-)}_\sigma \psi_{L \nu}, 
\ee
where $\tilde{R}{^{(-)ij}}_{\mu \nu}$ is the curvature 
of antiself-dual connection. 
The commutator algebra of the above supersymmetry 
on the complex tetrad, for example, is easily calculated as 
\ba
\A \A [\delta_{R1}, \delta_{R2}] e_{\mu}^i = 0 
= [\delta_{L1}, \delta_{L2}] e_{\mu}^i, \nonu
\A \A [\delta_{R1}, \delta_{L2}] e_{\mu}^i 
= 2i D_{\mu} (\overline{\tilde \alpha}_{2R} \gamma^i \alpha_{1R}). 
\ea
Therefore, we have 
\be
[\delta_1, \delta_2] e_{\mu}^i 
= 2i D_{\mu} (\overline{\tilde \alpha}_{2R} \gamma^i \alpha_{1R} 
  + \overline \alpha_{2L} \gamma^i \tilde \alpha_{1L}), 
\label{com-e}
\ee
where $\delta := \delta_R + \delta_L$. 
If the tetrad is real and $\psi_{R \mu} = \tilde \psi_{R \mu}$, 
the algebra of (\ref{com-e}) coincides with 
that of $N = 1$ supergravity \cite{FN}. 
The commutator algebra on (Majorana) Rarita-Schwinger fields 
is now being calculated. 

Finally we comment on the canonical formulation of ${\cal L}^{tot}$ 
in the source-free case. 
Although the Lagrangian ${\cal L}^{tot}$ involves 
the complex conjugate connection $\overline {A^{(+)}_{ij \mu}}$, 
the canonical formulation of ${\cal L}^{tot}$ 
in the source-free case 
can be expressed by using real canonical variables 
in a simple form. 
We assume spacetime manifold to be topologically $\Sigma \times R$, 
for some space-like submanifold $\Sigma$. 
Performing the Legendre transform, the ${\cal L}^{tot}_G$ of (\ref{LGtot}) 
can be written in canonical form: 
\be
{\cal L}^{tot}_G = - {\rm Tr}({\pi}^{(+)a} {\dot{A}}^{(+)}_a) 
        - {\rm Tr}(A^{(+)}_0 {\cal G}^{(+)}) 
        - N^a {\cal H}^{(+)}_a - N {\cal H}^{(+)} 
            + {\rm c.c.}, 
\label{LGtot-canc}
\ee
where the first four terms of right hand side of (\ref{LGtot-canc}) 
just comes from ${\cal L}^{(+)}$. 
Here $N^a$ is the complex shift vector, 
$N$ is a density of weight $-1$ representing the complex lapse, and 
\ba
{\cal G}^{(+)ij} \A := \A D^{(+)}_a {\pi}^{(+)ija} 
                     = \partial_a {\pi}^{(+)ija} 
                      + {[A^{(+)}_a, {\pi}^{(+)a}]}^{ij} \approx 0 
\label{cns+1} \\
{\cal H}^{(+)}_a \A := \A - {\rm Tr}({\pi}^{(+)b} R^{(+)}_{ab}) 
\approx 0 \\
{\cal H}^{(+)} \A := \A {\rm Tr}({\pi}^{(+)a} {\pi}^{(+)b} R^{(+)}_{ab}) 
\approx 0 
\label{cns+3}
\ea
are the Gauss, vector and Hamiltonian constraint, respectively, 
with ${\pi}^{(+)ija}$ the densitized momenta conjugate 
to $A^{(+)}_{ija}$.
\footnote{\ Latin letters {\it a, b,} $\cdots$ 
are spatial indices on $\Sigma$.} 
The trace in (\ref{LGtot-canc}) to (\ref{cns+3}) is defined as 
\be
{\rm Tr}(X Y) := {X_i}^j {Y_j}^k \ \ \ \ \ \ \ \ 
{\rm Tr}(X Y Z) := {X_i}^j {Y_j}^k {Z_k}^l. 
\ee
If we define real canonical variables $A_{ija}$ 
and ${\pi}^{ija}$ from the $A^{(+)}_{ija}$ 
and its conjugate momenta ${\pi}^{(+)ija}$ as 
\ba
A_{ija} \A := \A A^{(+)}_{ija} + \overline{A^{(+)}_{ija}}, \\
{\pi}^{ija} \A := \A {\pi}^{(+)ija} + \overline{{\pi}^{{(+)}ija}}, 
\ea
and split the $N^a$ and $N$ into the real and imaginary part 
as $N^a_1 + iN^a_2$ and $N_1 + iN_2$, 
then the ${\cal L}^{tot}_G$ of (\ref{LGtot-canc}) becomes 
\be
{\cal L}^{tot}_G = - {\rm Tr}({\pi}^{a} {\dot{A}}_a) 
        - {\rm Tr}(A_0 {\cal G}) 
        - N^a_1 {\cal H}_a - N^a_2 {\cal H}^*_a 
        - N_1 {\cal H} - N_2 {\cal H}^*, 
\label{LGtot-canr}
\ee
where 
\ba
{\cal G}^{ij} \A := \A {\cal G}^{(+)ij} + {\rm c.c.} = D_a {\pi}^{ija} 
                     = \partial_a {\pi}^{ija} 
                      + {[A_a, {\pi}^a]}^{ij} \approx 0 
\label{cns1} \\
{\cal H}_a \A := \A {\cal H}^{(+)}_a + {\rm c.c.} 
               = - {\rm Tr}({\pi}^b R_{ab}) \approx 0 \\
{\cal H}^*_a \A := \A i({\cal H}^{(+)}_a - {\rm c.c.}) 
                 = - {\rm Tr}({\pi}^b R^*_{ab}) \approx 0 \\
{\cal H} \A := \A {\cal H}^{(+)} + {\rm c.c.} 
             = {\rm Tr}({\pi}^a {\pi}^b R_{ab}) \approx 0 \\
{\cal H}^* \A := \A i({\cal H}^{(+)} - {\rm c.c.}) 
               = {\rm Tr}({\pi}^a {\pi}^b R^*_{ab}) \approx 0 
\label{cns5}
\ea
with the $\ast$ operation meaning the duality operation in Lorentz indices. 
These constraints of (\ref{cns1}) to (\ref{cns5}) 
form a first-class set, because the constraints of 
(\ref{cns+1}) to (\ref{cns+3}) are first-class. 
The canonical formulation of (\ref{LGtot-canr}) is just 
the $SO(3,1;R)$ theory which is derived 
from a viewpoint of generalization of the $SO(3;C)$ 
Ashtekar formulation \cite{PP}. 

In the source-free case, the canonical formulation 
of ${\cal L}^{tot}$ is equivalent to that of ${\cal L}^{(+)}$. 
When (Majorana) Rarita-Schwinger fields are coupled, 
however, the canonical formulation of ${\cal L}^{tot}$ 
may possibly differ from that of ${\cal L}^{(+)}$, 
because the term of (\ref{ImL+}) has canceled with 
its complex conjugate in ${\cal L}^{tot}$. 
The canonical formulation of ${\cal L}^{tot}$ 
including (Majorana) Rarita-Schwinger fields is under study.

We would like to thank the members of Physics Department 
at Saitama University 
for discussions and encouragement.



\end{document}